\documentclass[preprint,aps,showpacs,superscriptaddress]{revtex4}

\usepackage{graphicx}
\usepackage{dcolumn}
\usepackage{bm}

\begin{document}

\title{$\bar{K}NN$ quasi-bound state and the $\bar{K}N$
interaction: coupled-channel Faddeev calculations of the
$\bar{K}NN - \pi \Sigma N$ system}

\author{N.V. Shevchenko\footnote{Corresponding author: shevchenko@ujf.cas.cz}}
\affiliation{Nuclear Physics Institute, 25068 \v{R}e\v{z}, Czech Republic}

\author{A. Gal}\affiliation{Racah Institute of Physics,
The Hebrew University, Jerusalem 91904, Israel}

\author{J. Mare\v{s}}\affiliation{Nuclear Physics Institute,
25068 \v{R}e\v{z}, Czech Republic}

\author{J. R\'{e}vai}\affiliation{Research Institute for Particle and
Nuclear Physics, H-1525 Budapest, P.O.B. 49, Hungary}

\date{\today}

\begin{abstract}
Coupled-channel three-body calculations of an $I=1/2$,
$J^{\pi}=0^-$ $\bar{K}NN$ quasi-bound state in the $\bar{K}NN -
\pi \Sigma N$ system were performed and the dependence of the
resulting three-body energy on the two-body $\bar{K}N - \pi
\Sigma$ interaction was investigated. Earlier results of binding
energy $B_{K^-pp} \sim 50 -70$~MeV and width $\Gamma_{K^-pp} \sim
100$~MeV are confirmed [N.V. Shevchenko {\it et al.}, Phys. Rev.
Lett. {\bf 98}, 082301 (2007)]. It is shown that a suitably
constructed energy-independent complex $\bar{K}N$ potential gives
a considerably shallower and narrower three-body quasi-bound
state than the full coupled-channel calculation. Comparison with
other calculations is made.
\end{abstract}

\maketitle

\section{Introduction}

Hadronic nuclei are useful tools for studying hadron-nucleon interactions
and in-medium properties of hadrons. The recent interest in kaonic nuclei was
motivated by the strongly attractive antikaon-nucleus density-dependent
optical potentials obtained from $K^-$ atomic data fits~\cite{Batty}.
Akaishi and Yamazaki~\cite{Akaishi1} using G-matrix one-channel $\bar{K}N$
interactions, predicted the existence of deep and narrow $K^-$ bound states
in ${}^3$He, ${}^4$He, and ${}^8$Be. Of particular interest is the lightest
possible antikaon-nucleus system, $K^- p p$, for which these authors
calculated in Ref.~\cite{Akaishi2} values of 48 MeV and 61 MeV for the total
binding energy and the decay width, respectively.
Deeply bound kaonic states were searched in $^4$He$(K^-, N)$ reactions
at KEK, with negative results so far~\cite{Iwasaki}, and by the FINUDA
spectrometer collaboration at DA$\Phi$NE~\cite{FINUDA} in stopped $K^-$
reactions on nuclear targets such as lithium and carbon. The latter
experiment suggested evidence for a bound state $K^- p p$ `observed'
through its decay into approximately back-to-back $\Lambda$-proton pairs.
The deduced binding energy (115 MeV), but not the width (67 MeV), differs
considerably from the theoretical prediction of Ref.~\cite{Akaishi2}.
However, this interpretation of the measured $\Lambda$-proton spectrum
in the FINUDA experiment was challenged by Magas {\it et al.}~\cite{Oset}
who also criticized the Yamazaki-Akaishi calculations~\cite{Akaishi2}
for using an effective $\bar{K}$ optical potential in lieu of genuine
$\bar{K} N$ interactions.

The near-threshold $\bar{K}N$ interaction is mainly affected by
the sub-threshold $I=0$ resonance $\Lambda(1405)$, which is
usually assumed a $\bar{K}N$ bound state and a resonance in the
$\pi \Sigma$ channel. Numerous theoretical works were devoted to
constructing $\bar{K}N$ interactions within K-matrix models,
dispersion relations, meson-exchange models, quark models, cloudy
bag-models, and more recently by applying SU(3) meson-baryon
chiral perturbation theory (see e.g. the recent review
papers~\cite{Oller,Weise}). Scattering experiments for $K^- p$ are
rather old and the data are not too accurate. Kaonic hydrogen
provides additional information. Namely, there are two
experimental measurements of the $1s$ level shift and width caused
by the strong interaction, performed at KEK~\cite{KEK} and
recently by the DEAR collaboration at DA$\Phi$NE,
Frascati~\cite{DEAR}. The measured upward shift appears as due to
a repulsive strong interaction, but in fact it is caused by an
attractive interaction in the $I=0$ $\bar{K}N - \pi \Sigma$
channel, which is strong enough to generate a quasi-bound
strong-interaction state. The effect of such a strong attractive
interaction is to push the purely Coulomb level upwards. Using the
Deser formula~\cite{Deser}, it is possible to obtain the $K^- p$
scattering length from the value of the $1s$ level energy shift.
Unfortunately, several recent theoretical models could not
simultaneously reproduce the DEAR value of the $K^- p$ scattering
length together with the bulk of $K^-p$ scattering
data~\cite{Borasoy}.

As should be clear from this brief introduction, the fields of
$\bar{K}N$ and $\bar{K}$--nucleus interaction are abundant with
open questions and problems.
The elucidation of $\bar{K}$--
nuclear properties would help considerably to derive significant
information on the in-medium $\bar{K}N$ interaction
and on the possibility of kaon condensation in dense nuclear matter,
see Refs.~\cite{kondens1,kondens2} and previous works cited therein.
Among
$\bar{K}$-- nuclear systems, the study of three-body `exotic'
systems offers the advantage that Faddeev
equations~\cite{Faddeev}, which exactly describe the dynamics of
few particles, provide a proper theoretical and computational
framework. In the present work, we have generalized the Faddeev
equations in the Alt-Grassberger-Sandhas form~\cite{AGS} in order
to include additional `particle' channels and thus performed the
first genuinely three-body ${\bar K}NN - \pi \Sigma N$
coupled-channel Faddeev calculation in search for quasi-bound
states in the $K^- p p$ system. A preliminary report of this work
was given in Ref.~\cite{ourPRL}. The present paper provides a more
detailed and complete version of the previous one, especially
concerning the dependence of the three-body results on the
two-body input. The main result of Ref.~\cite{ourPRL} is
reconfirmed, namely that a single $K^-pp$ $I=1/2,~J^{\pi}=0^-$
quasi-bound state exists with binding energy $B \sim 50 - 70$~MeV
and width $\Gamma \sim 100$~MeV. It is shown that `equivalent'
single-channel ${\bar K}NN$ calculations of the type reported by
Yamazaki and Akaishi~\cite{Akaishi2} underestimate considerably
the binding energy, and particularly the width resulting within
the full ${\bar K}NN - \pi \Sigma N$ coupled-channel calculations.

The paper is organized as follows: in Section II we describe the derivation
of the coupled-channel Faddeev equations in the AGS form. The two-body
potentials which enter these equations are described in Section III.
Results are given in Section IV for the full coupled-channel calculations,
along with suitably chosen single-channel calculations that could provide
a testground for comparison with the single-channel calculation of
Ref.~\cite{Akaishi2}. Conclusions are given in Section V.

\section{Formalism}

Three-body Faddeev equations~\cite{Faddeev} in the
Alt-Grassberger-Sandhas (AGS) form~\cite{AGS}
\begin{equation}
\nonumber
U_{ij} = (1-\delta_{ij}) G_0^{-1}  + \sum_{k=1}^3 (1-\delta_{ik}) \,
T_k \, G_0 \, U_{kj} \\
\label{AGS}
\end{equation}
define unknown operators $U_{ij}$, describing the elastic and
re-arrangement processes $j + (ki) \to i + (jk)$. The inputs for
the AGS system of equations~(\ref{AGS}) are two-body $T$-matrices,
immersed into three-body space. The operator $G_0$ is the free three-body
Green's function. Faddeev partition indices $i,j = 1,2,3$ denote
simultaneously an interacting pair and a spectator particle.
When the initial state is known, as is usually
assumed, the system~(\ref{AGS}) consists of three equations.

The AGS equations are quantum-mechanical ones, describing
processes in which the number and composition of particles are
fixed. However, the two-body $\bar{K}N$ interaction, which is
essential for the $K^- pp$ quasi-bound state calculation, is
strongly coupled to other channels, particulary to the $\pi
\Sigma$ channel via $\Lambda(1405)$ . To take the $\bar{K}N - \pi
\Sigma$ coupling directly into account (we neglect the weaker coupled
$I=1$ $\pi \Lambda$ channel), it is necessary to extend the
formalism of Faddeev equations. To this end it is assumed that in
addition to the usual Faddeev channels, which represent different
partitions of the same set of particles, there are also `particle'
channels. Each of the three `particle' channels consists of three
usual Faddeev partitions (here we treat the two nucleons as
distinguishable particles, with proper antisymmetrization
introduced at a later stage). Thus, all three-body operators will
have `particle' indices ($\alpha$) for each state in addition to
the usual Faddeev indices ($i$), see Table~\ref{channels.tab}.
\begin{center}
\begin{table}[h]
\caption{Interacting two-body subsystems for three partition ($i$) and
three `particle' channel ($\alpha$) indices. The interactions are
further labelled by the two-body isospin values, entering the AGS
equations with total three-body isospin $I=1/2$.} \label{channels.tab}
\begin{tabular}{cccc}
\hline \hline
\quad $i$ $\setminus$ $\alpha$ \, & \, $1$ ($\bar{K}NN$)\, & \, $2$
($\pi \Sigma N$)\, & \, $3$ ($\pi N \Sigma$)\, \\[\smallskipamount]
\hline
\, 1 \, & $NN_{\, I=0,1}$ & $\Sigma N_{\, I=\frac{1}{2},\frac{3}{2}}$
    & $\Sigma N_{\, I=\frac{1}{2},\frac{3}{2}}$ \\
\, 2 \, & $\bar{K}N_{\, I=0,1}$ & $\pi N_{\, I=\frac{1}{2},\frac{3}{2}}$
    & $\pi \Sigma_{\, I=0,1}$\\
\, 3 \, & $\bar{K}N_{\, I=0,1}$ & $\pi \Sigma_{\, I=0,1}$
    & $\pi N_{\, I=\frac{1}{2},\frac{3}{2}}$ \\[\smallskipamount]
\hline \hline
\end{tabular}
\end{table}
\end{center}

All operators in Eq.~(\ref{AGS}) now act in this additional
`particle' space: $T_i$ transform to $T_i^{\alpha \beta}$, $G_0
\to G_0^{\alpha \beta}$, and $U_{ij} \to U_{ij}^{\alpha \beta}$
($\alpha, \beta = 1, 2, 3$). The two-body $T$-matrices have the
following form:
\begin{equation}
\label{T3x3}
 T_1 \to  \left(
    \begin{tabular}{ccc}
    $T_1^{NN}$ & 0 & 0 \\
    0 & $T_1^{\Sigma N}$ & 0 \\
    0 & 0 & $T_1^{\Sigma N}$
    \end{tabular}
\right), \quad
 T_2 \to  \left(
    \begin{tabular}{ccc}
    $T_2^{KK}$ & 0 & $T_2^{K \pi}$ \\
    0 & $T_2^{\pi N}$ & 0 \\
    $T_2^{\pi K}$ & 0 & $T_2^{\pi \pi}$
    \end{tabular}
\right), \quad
 T_3 \to  \left(
    \begin{tabular}{ccc}
    $T_3^{KK}$ & $T_3^{K \pi}$ & 0 \\
    $T_3^{\pi K}$ & $T_3^{\pi \pi}$ & 0 \\
    0 & 0 & $T_3^{\pi N}$
    \end{tabular}
\right) \,,
\end{equation}
where $T_i^{NN}$, $T_i^{\pi N}$ and $T_i^{\Sigma N}$ are the usual
one-channel two-body $T$-matrices in three-body space, describing
$NN$, $\pi N$, and $\Sigma N$ interactions, respectively.
The elements of the coupled-channel $T$-matrix,
$T_i^{KK}$, $T_i^{\pi \pi}$, $T_i^{\pi K}$, and $T_i^{K \pi}$, are
labelled by two meson indices:
\begin{eqnarray*}
T_i^{KK}:&      \qquad \bar{K} + N & \to \bar{K} + N \\
T_i^{\pi K}:&   \qquad \bar{K} + N & \to \pi + \Sigma \\
T_i^{K \pi}:&   \qquad \pi + \Sigma & \to \bar{K} + N \\
T_i^{\pi \pi}:& \qquad \pi + \Sigma & \to \pi + \Sigma ~.
\end{eqnarray*}
The free Green's function is diagonal in channel indices:
$G_0^{\alpha \beta} = \delta_{\alpha \beta} \, G_0^{\alpha}$,
while the transition operators $U_{ij}^{\alpha \beta}$ have the
most general form.

Searching for quasi-bound states assumes working at low energies.
Low-energy interactions are satisfactorily described by $s$-waves,
hence for all the relevant two-body interactions we use $L_i=0$.
The total orbital angular momentum is then $L=0$. For the $K^- pp$
system, the total spin is $S=0$ due to the spin zero of the two protons
and spin zero of the $K^-$ meson. All two-body baryon-baryon
interactions are then spin-zero interactions. The remaining quantum number
is isospin. It is possible to work in either particle or isospin basis, but
since the Coulomb interaction is not included in the present calculation and
charge independence is assumed for all two-body interactions, it is quite
natural to choose the isospin basis.
The total isospin $I$ is a conserved quantum number for charge-independent
interactions, so a bound (or a quasi-bound) state must have a definite value
of $I$. For $I=1/2$
there are two possible (unadmixed) states corresponding to the total
spin $S$ of the system. In the $\bar{K}NN - \pi \Sigma N$ case $S$ coincides with
the spin of the two baryons ($S_i=0,1$) and due to their indistinguishability
the spin value also fixes the isospin of the two nucleons, $I_{NN}=1,0$,
respectively. In these states --
let us call them $pp$- and $d$-configuration -- a more attractive combination
of $\bar{K}N$ $I=0,1$
forces and a weaker $NN$ singlet force in the $pp$ is competing with a weaker
$\bar{K}N$ attraction and a stronger $NN$ triplet force in $d$. Therefore it
is not clear a priori, which of them has a lower energy.
We have chosen to calculate the $I=1/2$, $S=0$ $pp$
configuration due to its connection to experiment. Moreover,
simple isospin re-coupling arguments indicate, that it might have a lower energy.
However, a similar calculation should be
performed for the other, $I=1/2$, $S=1$ $d$-configuration, too.
As for the $I=3/2$ state, it is governed by a weaker $\bar{K}N$
attraction than the one in the $I=1/2$ state under
consideration in this work.

Separable potentials, and the corresponding $T$-matrices, are widely used
in Faddeev calculations for reducing the dimension of integrals in the
equations. The separable-potential approximation is justified
by the fact that the kernels of two-particle equations are of the
Hilbert-Schmidt type, at least under suitable conditions on the
two-particle interactions~\cite{Meetz}. Namely, the separable approximation
is valid when each of the two-particle subsystems is dominated by a limited
number of bound states or resonances~\cite{Lovelace}. This condition is
satisfied for the `main' two-body interactions entering our system,
$\bar{K}N-\pi \Sigma$ and $NN$. For the remaining $\Sigma N$ and $\pi N$
interactions we expect weaker contributions to the bound-state complex
energy (as already demonstrated for $\Sigma N$ in Ref.~\cite{ourPRL}).
Hence we use for all two-body potentials a separable form:
\begin{equation}
\label{Voperator}
 V_{i,I}^{\alpha \beta} = \lambda_{i,I}^{\alpha \beta} \,
 |g_{i,I}^{\alpha} \rangle  \langle g_{i,I}^{\beta} | \,,
\end{equation}
which leads to a separable form of $T$-matrices:
\begin{equation}
\label{Toperator}
 T_{i,I}^{\alpha \beta} = |g_{i,I}^{\alpha} \rangle
\tau_{i,I}^{\alpha \beta} \langle g_{i,I}^{\beta} | \,.
\end{equation}
For $\alpha = \beta$ the corresponding $T$-matrix coincides with
the usual one. With the relation~(\ref{Toperator}), the AGS
system~(\ref{AGS}) can be expressed using new
transition and kernel operators:
\begin{eqnarray}
\label{X_definition}
X_{ij, I_i I_j}^{\alpha \beta} &=&  \langle
g_{i,I_i}^{\alpha} | G_0^{\alpha} \, U_{ij, I_i I_j}^{\alpha
\beta} G_0^{\beta} | g_{j,I_j}^{\beta} \rangle \,, \\
\label{Z_definition}
Z_{ij, I_i I_j}^{\alpha \beta} &=&
\delta_{\alpha \beta} \, Z_{ij, I_i I_j}^{\alpha} =
\delta_{\alpha \beta} \, (1-\delta_{ij}) \,
\langle g_{i,I_i}^{\alpha} | G_0^{\alpha} | g_{j,I_j}^{\alpha}
\rangle \,.
\end{eqnarray}
Substituting isospin-dependent $T_i^{\alpha \beta}$, $Z_{ij}^{\alpha}$,
and $X_{ij}^{\alpha \beta}$ into the AGS system~(\ref{AGS}) we
obtain the following system of operator equations:
\begin{equation}
\label{full_oper_eq}
X_{ij, I_i I_j}^{\alpha \beta} = \delta_{\alpha \beta} \,
Z_{ij, I_i I_j}^{\alpha} +
\sum_{k=1}^3 \sum_{\gamma=1}^3 \sum_{I_k}
Z_{ik, I_i I_k}^{\alpha} \, \tau_{k, I_k}^{\alpha \gamma} \,
X_{kj, I_k I_j}^{\gamma \beta} \,.
\end{equation}
The number of equations in the system is defined by the number
of possible form-factors $g_{i, I_i}^{\alpha}$. As
is seen from Table~\ref{channels.tab}, before antisymmetrization
our system consists of 18 equations.

Three sets of Jacobi momentum coordinates should be introduced for
each `particle' channel $\alpha$: $| \vec{k_{i}}^{\alpha},
\vec{p_i}^{\alpha} \rangle$, $i=1,2,3$,
$\alpha=1,2,3$. Here, $\vec{k_{i}}^{\alpha}$ is the
center-of-mass momentum of the $(jk)$ pair and $\vec{p_{i}}^{\alpha}$
is the momentum
of spectator $i$ with respect to the pair $(jk)$, $i \neq j \neq k$.
In these coordinates the three-body free Hamiltonian in the
channel $\alpha$ is defined as
\begin{equation}
H_0^{\alpha} = \frac{(k_{i}^{\alpha})^2}{2 \, m_{jk}^{\alpha}} +
\frac{(p_{i}^{\alpha})^2}{2 \, \mu_{i}^{\alpha}} \,,
\end{equation}
where the reduced masses also have `particle' channel indices:
\begin{equation}
m_{jk}^{\alpha} = \frac{m_j^{\alpha} m_k^{\alpha}}
{m_j^{\alpha} + m_k^{\alpha}}, \quad
\mu_{i}^{\alpha} = \frac{m_i^{\alpha} (m_j^{\alpha} + m_k^{\alpha})}
{m_i^{\alpha} + m_j^{\alpha} + m_k^{\alpha}}, \quad i \neq j \neq k \,.
\end{equation}
In contrast to the usual AGS formalism we have to use not the kinetic energy,
but the total energy of the system, including rest masses. We introduce
threshold energies: $z_{th}^{\alpha} = \sum_{i=1}^3 m_i^{\alpha}$,
so that the total energy is $z_{tot} = z_{th}^{\alpha} + z_{kin}^{\alpha}$,
where $z_{kin}^{\alpha}$ denotes the kinetic energy in channel $\alpha$.
The integrations in Eqs.~(\ref{X_definition}) and (\ref{Z_definition})
are performed over one of the Jacobi momenta, namely, over $\vec{k_{i}}^{\alpha}$,
which describes the motion of an interacting pair of particles $j$ and $k$
($i \neq j \neq k$). Thus, the operators $X$ and $Z$ act on the second
momentum, $\vec{p_{i}^{\alpha}}$:
\begin{eqnarray}
\label{X_mom}
&{}& X_{ij, I_i I_j}^{\alpha \beta} \to
\left\langle \vec{p_i}^{\alpha} | X_{ij, I_i I_j}^{\alpha \beta}
(z_{tot}) | \vec{p_j}'^{\beta} \right \rangle
=  X_{ij, I_i I_j}^{\alpha \beta}
(\vec{p_i}^{\alpha}, \vec{p_j}'^{\beta}; z_{kin}^{\alpha} + z_{th}^{\alpha}) \,, \\
\label{Z_mom}
&{}& Z_{ij, I_i I_j}^{\alpha} \to
\left\langle \vec{p_i}^{\alpha} | Z_{ij, I_i I_j}^{\alpha}
(z_{tot}) |
\vec{p_j}'^{\alpha} \right \rangle
=  Z_{ij, I_i I_j}^{\alpha}
(\vec{p_i}^{\alpha}, \vec{p_j}'^{\alpha}; z_{kin}^{\alpha} + z_{th}^{\alpha})
\,.
\end{eqnarray}
The energy-dependent part of a two-body $T$-matrix, embedded in the
three-body space is defined by the following relation:
\begin{equation}
\label{tau_mom}
\tau_{i,I_i}^{\alpha \beta} \to
\left\langle \vec{p_i}^{\alpha} | \tau_{i, I_i}^{\alpha \beta}(z_{tot}) |
\vec{p_j}'^{\beta} \right \rangle  \equiv
\delta_{ij} \, \delta(\vec{p_i}^{\alpha} - \vec{p_j}'^{\beta}) \,
\tau_{i,I_i}^{\alpha \beta}\left(
z_{tot} - z_{th}^{\alpha} - \frac{(p_i^{\alpha})^2}{2 \, \mu_i}
\right) \,.
\end{equation}
It is worth noting that all elements of the two-channel
two-body $\bar{K}N - \pi \Sigma$ $T$-matrix depend on
the kinetic energies in both channels ($z_{kin}^1$ and $z_{kin}^2$)
simultaneously.
Here we define the argument of the corresponding $\tau^{\alpha \beta}$
using the left `particle' index $\alpha$. The second kinetic energy can be simply
found from the relation
$z_{kin}^{\alpha} + z_{th}^{\alpha} = z_{kin}^{\beta} + z_{th}^{\beta}$.

The calculation of the kernels $Z$ involves transformation from
one set of Jacobi coordinates to another one and isospin
re-coupling, using the property of free Green's function:
\begin{equation}
\left\langle \vec{p_i}^{\alpha}, I_i^{\alpha} | G_0^{\alpha} |
\vec{p_j}'^{\alpha}, I_j^{\alpha} \right\rangle
= \left\langle \vec{p_i}^{\alpha} | G_0^{\alpha} | \vec{p_j}'^{\alpha}
\right\rangle_{I_i^{\alpha} I_j^{\alpha}} \,
\left\langle  i_j^{\alpha} \, i_k^{\alpha}  (I_i^{\alpha}) \, i_i^{\alpha},
I I_z | i_i^{\alpha} \, i_k^{\alpha}  (I_j^{\alpha}) \, i_j^{\alpha},
I I_z \right\rangle \,,
\end{equation}
where $i_j^{\alpha}$ and $I_j^{\alpha}$ denote one-particle and two-particle
isospins, respectively, with partition subscripts $i \neq j \neq k$,
the total three-body isospin and its projection being $I=1/2, I_z = 1/2$.

To search for a resonance or a bound state means to look for
a solution of the homogeneous system corresponding to Eq.~(\ref{full_oper_eq}).
But before solving the system
\begin{equation}
\label{homog_oper_eq}
X_{i, I_i}^{\alpha} =
\sum_{k=1}^3 \sum_{\gamma=1}^3 \sum_{I_k}
Z_{ik, I_i I_k}^{\alpha} \, \tau_{k, I_k}^{\alpha \gamma} \,
X_{k, I_k}^{\gamma} \,,
\end{equation}
we must antisymmetrize operators involving two identical
baryons with antisymmetric spin components ($S_i=0$) and symmetric spatial
components ($L_i=0$). Here, in Eq.~(\ref{homog_oper_eq}), and in the following
we omit right-hand indices of $X$:
$X_{ij,I_i I_j}^{\alpha \beta} \to X_{i,I_i}^{\alpha}$,
which are unnecessary for a homogeneous system.
The operator $X_{1,0}^1$ has antisymmetric $NN$ isospin components,
so it drops out of the equations. In contrast, the operator
$X_{1,1}^1$ has the correct symmetry properties. All the remaining
operators form symmetric and antisymmetric pairs, the symmetric
ones which are used in the calculation are:
\begin{eqnarray}
\nonumber
&{}&X_{2,0}^{1,-} = X_{2,0}^{1} - X_{3,0}^{1}, \quad
 X_{2,1}^{1,+} = X_{2,1}^{1} + X_{3,1}^{1}, \\
\nonumber
&{}&X_{2,0}^{3,-} = X_{2,0}^{3} - X_{3,0}^{2}, \quad
 X_{2,1}^{3,+} = X_{2,1}^{3} + X_{3,1}^{2}, \\
\label{X_symmetrical}
&{}&X_{1,\frac{3}{2}}^{2,-} = X_{1,\frac{3}{2}}^{2} - X_{1,\frac{3}{2}}^{3},
\quad
 X_{1,\frac{1}{2}}^{2,+} = X_{1,\frac{1}{2}}^{2} + X_{1,\frac{1}{2}}^{3}, \\
\nonumber
&{}&X_{2,\frac{3}{2}}^{2,-} = X_{2,\frac{3}{2}}^{2} - X_{3,\frac{3}{2}}^{3},
\quad
 X_{2,\frac{1}{2}}^{2,+} = X_{2,\frac{1}{2}}^{2} + X_{3,\frac{1}{2}}^{3} \,.
\end{eqnarray}
Taking into account equalities of some kernel functions,
we end up with a system of nine coupled operator equations in the eight
new operators~(\ref{X_symmetrical}) and $X_{1,1}^1$, all of which have
the required symmetry properties. Since the Faddeev equations are dynamical
ones, their final number after antisymmetrization corresponds to the
number of different form-factors entering the interactions. Similar
antisymmetrization procedures have been implemented in several
multi-channel Faddeev calculations, e.g. the fairly recent $K^- d$
work of Ref.\cite{Bahaoui}.

To solve the homogeneous system we
transform the integral equations into algebraic ones and then
search for the complex energy at which the determinant of the kernel
matrix becomes equal to zero. We are looking for a three-body pole,
the real part of which is situated between the $\bar{K}NN$ and
$\pi \Sigma N$ thresholds,
corresponding to a resonance in the $\pi \Sigma N$ channel and
a quasi-bound state (a bound state with non-zero width) in the
$\bar{K}NN$ channel. Therefore, we must work on the physical energy
sheet of channel one and on an unphysical sheet of the second
channel.

\section{Input}

The separable potential~(\ref{Voperator}),
in momentum representation, has a form:
\begin{equation}
\label{Vseprb}
 V_{i,I_i}^{\alpha \beta}(k_i^{\alpha},k_i'^{\beta}) =
 \lambda_{i,I_i}^{\alpha \beta} \;
 g_{i,I_i}^{\alpha}(k_i^{\alpha}) \, g_{i,I_i}^{\beta}(k_i'^{\beta}).
\end{equation}
For the $NN$, $\Sigma N$ and $\pi N$ interactions we have $\alpha = \beta$,
whereas for the coupled-channel $\bar{K}N - \pi \Sigma$ interaction
$\alpha, \beta = K$ ($\bar{K}N$-channel) or $\pi$
($\pi \Sigma$-channel). We constructed our own coupled-channel
$\bar{K}N - \pi \Sigma$ interactions, plus complex and real one-channel
$\bar{K}N$ test potentials discussed below. We also constructed one-channel
$\Sigma N$ interaction and used the PEST $NN$ potential~\cite{NNpot}.
Here we neglect the $\pi N$ interaction since its dominant part
is in the (3,3) $p$-wave channel.

\subsection{$\bar{K} N$ interaction}

\subsubsection{Two-channel $\bar{K}N - \pi \Sigma$}

\begin{figure}
\includegraphics[scale=0.4,angle=-90]{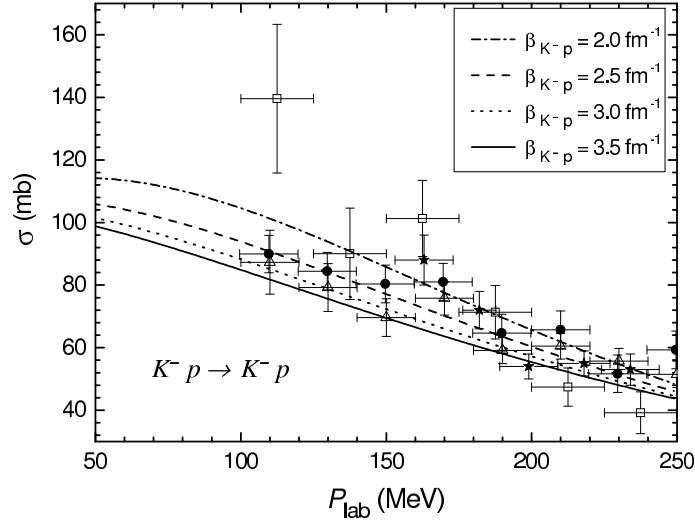}
\caption{\label{KpKpcross.fig} Total $K^- p \to K^- p$ cross
sections calculated for four sets of $\bar{K}N - \pi \Sigma$ parameters
with different values of $\beta$ marked in the inset.
The experimental values are taken from~\cite{Kp1exp} (open
squares), ~\cite{Kp2exp} (open triangles), ~\cite{Kp3exp} (solid
circles), and ~\cite{Kp4exp} (stars).}
\end{figure}

\begin{figure}
\includegraphics[scale=0.4,angle=-90]{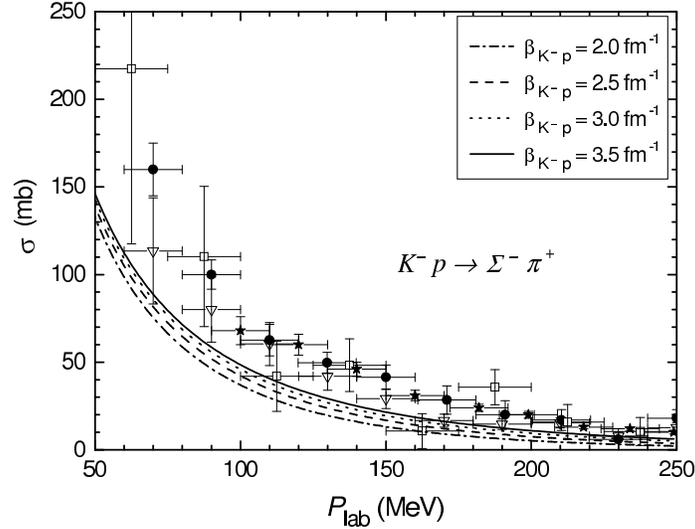}
\caption{\label{KppiSigcross.fig} Total $K^- p \to \pi^+ \Sigma^-$
cross sections calculated for four sets of $\bar{K}N - \pi \Sigma$
parameters with different values of $\beta$ marked in the inset.
The experimental values are taken from~\cite{Kp1exp} (open squares),
~\cite{Kp2exp} (open triangles), ~\cite{Kp3exp} (solid circles),
and ~\cite{Kp4exp} (stars).}
\end{figure}

There are many models of strangeness $-1$ meson-baryon scattering,
constructed using different methods, see e.g. Refs.~\cite{Borasoy,Oller2} and
references therein. These recent papers describe coupled-channel models of
the  $\bar{K}N$ interaction, constructed within the framework of Chiral
perturbation theory. The exclusive use of on-shell amplitudes and the amount
of coupled channels involved in such works
renders them impractical for Faddeev calculations. We therefore constructed
our own potentials for the coupled-channel $\bar{K}N - \pi \Sigma$
interaction in the form~(\ref{Vseprb}) with form-factors
\begin{equation}
g_{I}^{\alpha}(k^{\alpha}) = \frac{1}{(k^{\alpha})^2 +
(\beta_{I}^{\alpha})^2} \, .
\label{formfactorKN}
\end{equation}
To obtain the parameters $\lambda_{I}^{\alpha \beta}$ and
$\beta_{I}^{\alpha}$ we used the following experimental data:
\begin{enumerate}

\item[(i)] Mass $M_{\Lambda}$ and width $\Gamma_{\Lambda}$ of the $\Lambda(1405)$
resonance, assuming that it is a quasi-bound state in the $I=0$ $\bar{K}N$
channel and a resonance in the $I=0$ $\pi \Sigma$ channel. For the energy of
$\Lambda(1405)$ $E_{\Lambda} = M_{\Lambda} - {\rm i}~\Gamma_{\Lambda}/2$,
($c=\hbar=1$), we adopted the PDG value~\cite{PDG}
$E_{\Lambda}^{\, \rm PDG} = 1406.5 - {\rm i}~25$ MeV.
In some cases we used also other values of $M_{\Lambda}$ and $\Gamma_{\Lambda}$.

\item[(ii)] The $K^- p$ scattering length as derived from the atomic $1s$ level
shift and width in the KEK experiment~\cite{KEK}
\begin{equation}
a_{K^-p} = (-0.78 \pm 0.15 \pm 0.03) + {\rm i}~(0.49 \pm 0.25 \pm 0.12)
\; {\rm fm}
\end{equation}
and in the DEAR collaboration experiment~\cite{DEAR}
\begin{equation}
a_{K^-p} = (-0.468 \pm 0.090 \pm 0.015) + {\rm i}~(0.302 \pm 0.135 \pm
0.036) \; {\rm fm} \,.
\end{equation}
In the following we denote the KEK value as
$a_{K^-p}^{\rm KEK} = -0.78 + {\rm i}~0.49$ fm and the DEAR value
as $a_{K^-p}^{\rm DEAR} = -0.468 + {\rm i}~0.302$ fm.
Due to the fairly large experimental errors and also the large
difference between the results of these two measurements, we fitted our
parameters to a variety of values for the $K^- p$ scattering length.
In Ref.~\cite{ourPRL} we studied the sensitivity of the Faddeev calculations'
results to varying the KEK value within its error bars.
The three-body pole energy was found to depend strongly on the input $K^- p$
scattering length. As for the DEAR value of the $K^- p$ scattering length,
we note the controversy about its consistency with the bulk of the $K^- p$
scattering data~\cite{Borasoy,Oller2}.

\item[(iii)] The very accurately measured threshold branching ratio~\cite{gammaKp}:
\begin{equation}
\gamma = \frac{\Gamma(K^- p \to \pi^+ \Sigma^-)}{\Gamma(K^- p \to
\pi^- \Sigma^+)} = 2.36 \pm 0.04 \, .
\end{equation}
The value $2.36$ was used in our fits.

\item[(iv)] Elastic $K^- p \to K^- p$ and inelastic $K^- p \to \pi^+
\Sigma^-$ total cross sections. We chose these two reactions
because among all available cross section data they have sufficient
experimental data points with reasonable experimental errors.

\end{enumerate}
We fitted the potential parameters to points (i)--(iii) of this
list and then checked how well the resulting potential reproduces
the cross sections (iv). The calculated cross-sections for four sets
of parameters, in comparison with the experimental data, are shown in
Figs.~\ref{KpKpcross.fig} and~\ref{KppiSigcross.fig}. These sets differ
from each other by the value of the range parameter $\beta$;
the remaining parameters were also changed in order to
reproduce the same $\gamma$, $a_{K^- p}^{\rm KEK}$
and $E_{\Lambda}^{\, \rm PDG}$ data.
We conclude from the figures that the best value of the $\bar{K}N$
range parameter is $\beta=3.5$ fm$^{-1}$. In the following we denote
the set with $a_{K^- p}^{\rm KEK}$, $E_{\Lambda}^{\, \rm PDG}$, and
$\beta=3.5$ fm$^{-1}$ as the `best set'.


\begin{figure}
\includegraphics[scale=0.4,angle=-90]{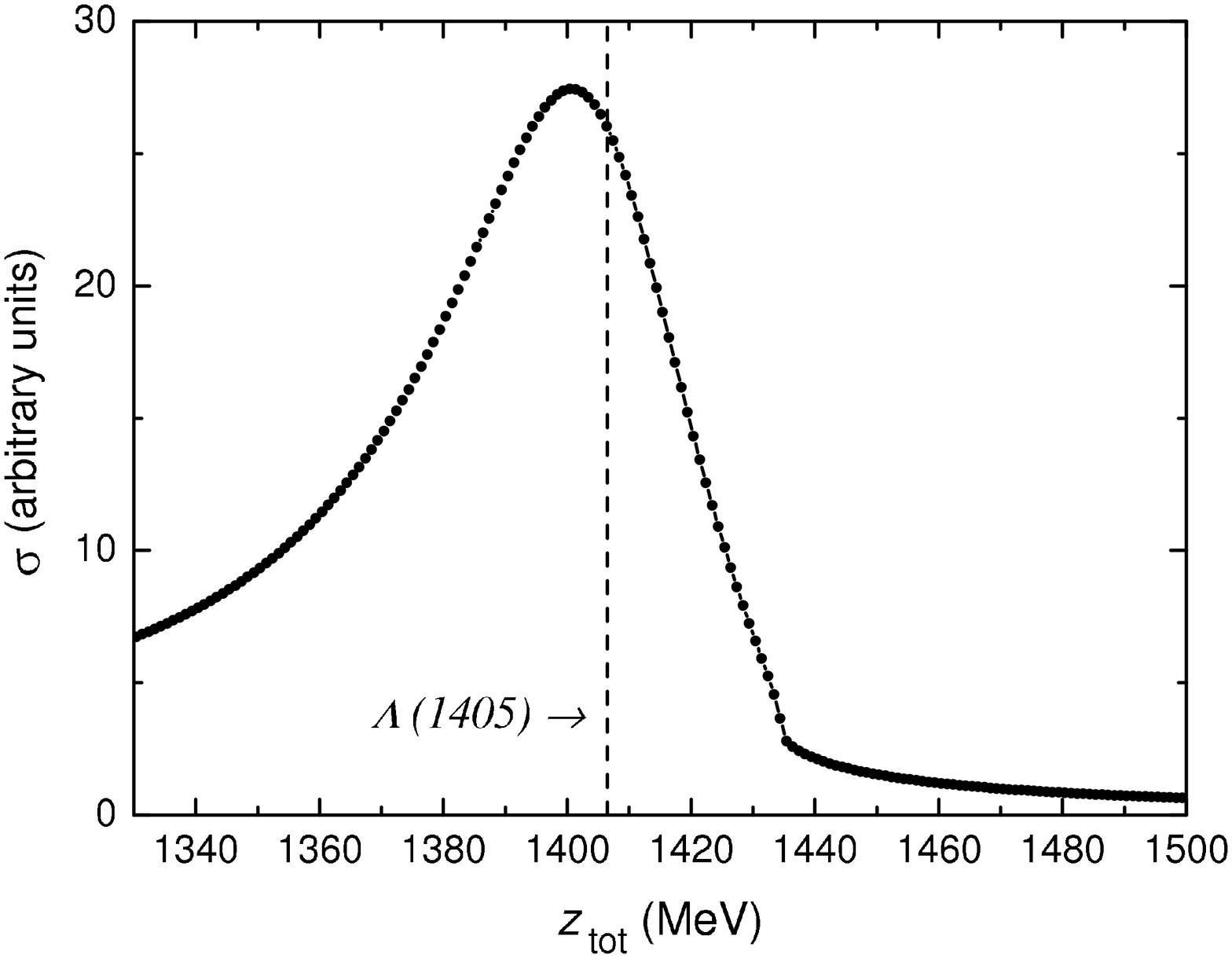}
\caption{\label{Sigmapi_I0.fig} Calculated elastic $\pi \Sigma$ cross
section for $I=0$, arbitrary units.}
\end{figure}

Figure~\ref{Sigmapi_I0.fig} shows the calculated $I=0$ elastic $\pi \Sigma$
cross section, demonstrating that $\Lambda(1405)$ is indeed a resonance
in this channel.

We were unable to find a value for $\beta$, using the DEAR
scattering length $a_{K^- p}^{\rm DEAR}$ and $E_{\Lambda}^{\, \rm PDG}$,
such that the corresponding set of parameters provided a good
description of both cross-sections.
The elastic $K^- p \to K^- p$ cross-sections can be described with
$1.5 \leq \beta \leq 2.5$~fm$^{-1}$, but the inelastic $K^- p \to
\pi^+ \Sigma^-$ cross sections for these values are situated much lower
than the experimental data points. Given this situation, we did not perform
three-body calculations with $\bar{K}N$ interaction parameters that
reproduce the DEAR value of the $K^- p$ scattering length.

\subsubsection{One-channel complex and real $\bar{K}N$}

In order to investigate all possible dependencies of our three-body
results on two-body inputs we constructed additionally real and complex
one-channel $\bar{K}N$ potentials. The imaginary part of the complex
potential accounts for absorption to all other channels.
Both potentials have the same form-factors as
the coupled-channel potential~[Eq.(\ref{formfactorKN})], but for only one channel
index $\alpha = \beta = K$. To fit the strength
parameters $\lambda$ of the complex variant, we used experimental data
(i) and (ii), i.e. the energy of $\Lambda(1405)$ and $a_{K^- p}$.
For the complex $\bar{K}N$ potential we used `best set' plus one more
set of data, which is the same as was used in Refs.~\cite{Akaishi1,Akaishi2}:
$E_{\Lambda}^{\,\rm AY} = 1405 - {\rm i}~20$ MeV,
$a_{K^- p}^{\,\rm AY} = -0.70 + {\rm i}~ 0.53$ fm, and a range parameter
$\beta=1.5$ fm$^{-1}$. We denote it as `AY set'.

A one-channel real $\bar{K}N$ potential was constructed by fitting its
parameters to reproduce the real parts of $E_{\Lambda}^{\, \rm PDG}$ and
$a_{K^- p}^{\rm KEK}$, with $\beta=3.5$ fm$^{-1}$. Here we assumed that
$\Lambda(1405)$ is a real bound state of the $I=0$ $\bar{K}N$ subsystem.

\subsection{$\Sigma N$ interaction}

Only few experimental data exist for this interaction. There are
different models of it, for example several Nijmegen models, but
due to the lack of data it is not possible to give preference to
any of these over the other ones. A separable potential~(\ref{Vseprb})
with Yamaguchi form-factors
\begin{equation}
g_{I}^{\Sigma N}(k) = \frac{1}{k^2 + (\beta_{I}^{\Sigma N})^2}
\end{equation}
was used for the two isospin states. The parameters of the $I=3/2$
$\Sigma N$ interaction were fitted to:
\begin{enumerate}
\item[(i)]  the scattering length and effective radius
\begin{equation}
a(I=3/2) = 3.8 \; {\rm fm}, \qquad r_{{\rm eff}}(I=3/2) = 4.0 \; {\rm fm}
\end{equation}
from the Nijmegen potential model F~\cite{SigmaN1} (we denote this set
of $I=3/2$ $\Sigma N$ parameters as '$\Sigma N$ set 1').

\item[(ii)] the Nijmegen model NSC97 $YN$ phase shifts~\cite{SigmaN2}. This
'$\Sigma N$ set 2' gives the following scattering length and effective range
\begin{equation}
a(I=3/2) = 4.15 \; {\rm fm}, \qquad r_{{\rm eff}}(I=3/2) = 2.4 \; {\rm fm.}
\end{equation}

\item[(iii)]  the scattering length and effective radius
\begin{equation}
a(I=3/2) = 4.1 \; {\rm fm}, \qquad r_{{\rm eff}}(I=3/2) = 3.5 \; {\rm fm}
\end{equation}
from the most recent Nijmegen potential ESC04a~\cite{SigmaN3} ('$\Sigma N$ set 3').

\end{enumerate}

\begin{center}
\begin{table}[hb]
\caption{Three-body pole
energy $E_{K^- pp}$ (in MeV) of the $I=1/2,~J^{\pi}=0^-$ quasi-bound
state of the $\bar{K} NN$ system with respect to the $K^-pp$ threshold
calculated
with the 'best set' of $\bar{K}N - \pi \Sigma$ parameters using
'$\Sigma N$ set 1', '$\Sigma N$ set 2', '$\Sigma N$ set 3', and with
both $I=1/2$ and $I=3/2$ $\Sigma N$ interactions switched off.}
\label{SigmaN.tab}
\begin{tabular}{cccc}
\hline \hline
'$\Sigma N$ set 1' & '$\Sigma N$ set 2' & '$\Sigma N$ set 3' &
no $\Sigma N$ \\
\hline
\, $-55.1-{\rm i}~50.9$ \, & \,  $-55.4-{\rm i}~51.9$ \, &
\,  $-55.3-{\rm i}~51.1$ \, & \, $-52.9-{\rm i}~50.9$ \\
\hline \hline
\end{tabular}
\end{table}
\end{center}
The dependence of the three-body pole position on the
$\Sigma N$ parameters was investigated in Ref.~\cite{ourPRL}.
Table~\ref{SigmaN.tab} illustrates the sensitivity of the
binding energies and widths of the $I=1/2,~J^{\pi}=0^-$ quasi-bound
state of the $\bar{K} NN$ system to the $\Sigma N$ interaction
parameters.
Due to the weak dependence of the three-body pole position on the
$\Sigma N$ interaction we used in the following only one (the first) set of
$I=3/2$ $\Sigma N$ parameters.

For the $I=1/2$ $\Sigma N$ interaction only the scattering length was
approximately determined: $a(I=1/2) = -0.5$ fm~\cite{DalitzSigmaN}.
We fitted the separable-potential parameters to this value, restricting
the fit by imposing `natural' values on the parameters and producing
a reasonable value for the $I=1/2$ effective radius.

\subsection{$NN$ interaction}

We used the nucleon-nucleon PEST potential from Ref.~\cite{NNpot},
which is a separable approximation of the Paris potential. The
strength parameter was set to $\lambda=-1$ and the form-factor is:
\begin{equation}
g_{I}^{NN}(k) = \frac{1}{2 \sqrt{\pi}} \, \sum_{i=1}^6
\frac{c_{i,I}^{NN}}{k^2 + (\beta_{i,I}^{NN})^2} \,.
\end{equation}
The constants $c_{i,I}^{NN}$ and $\beta_{i,I}^{NN}$ are listed
in Ref.~\cite{NNpot}. PEST is on- and off-shell equivalent to the Paris
potential up to $E_{\,\rm lab} \sim 50$ MeV and is repulsive at
distances shorter than 0.8 fm. It reproduces the deuteron binding
energy $E_{\,\rm d} = -2.2249$ MeV, as well as the triplet and singlet $NN$
scattering lengths, $a(\,{}^3S_1) = -5.422$ fm and $a(\,{}^1S_0) = 17.534$ fm,
respectively.

\section{Results}

\subsection{Results of full coupled-channel $\bar{K}NN - \pi \Sigma N$
calculation}

\begin{figure}[h]
\includegraphics[scale=0.4,angle=-90]{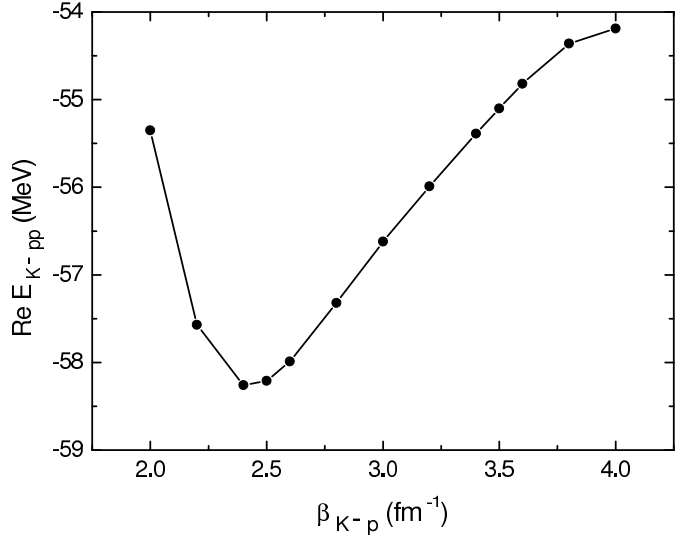}
\caption{\label{Re_diffbetas.fig} Coupled-channel calculation:
the real part of the three-body $\bar{K}NN - \pi \Sigma N$
pole energy as function of the $\bar{K}N$ range parameter $\beta$. The two-body
$\bar{K}N - \pi \Sigma$ observables are fixed at $a_{K^- p}^{\rm KEK}$
and $E_{\Lambda}^{\, \rm PDG}$.}
\end{figure}

\begin{figure}[h]
\includegraphics[scale=0.4,angle=-90]{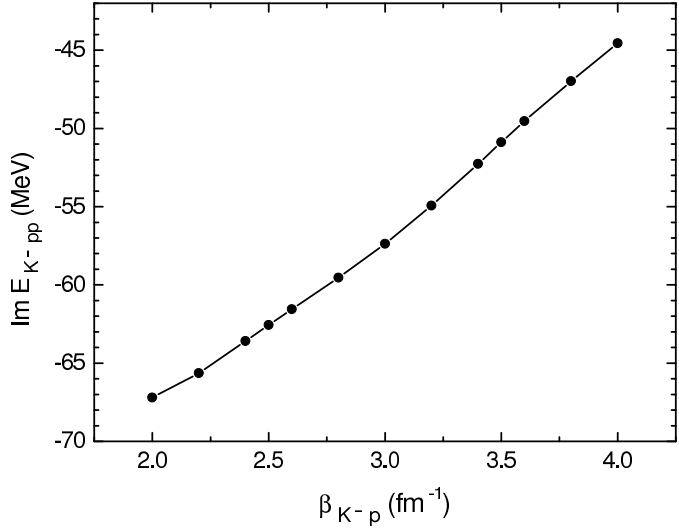}
\caption{\label{Im_diffbetas.fig} Coupled-channel calculation:
the imaginary part of the three-body $\bar{K}NN - \pi \Sigma N$
pole energy as function of the $\bar{K}N$ range parameter
$\beta$. The two-body $\bar{K}N - \pi \Sigma$ observables are fixed at
$a_{K^- p}^{\rm KEK}$ and $E_{\Lambda}^{\, \rm PDG}$.}
\end{figure}

Full coupled-channel calculations were done systematically, studying
various dependencies of the three-body pole position on different input
parameters of the $\bar{K}N - \pi \Sigma$ potential.
Here the three-body energy is defined as $E_{K^- pp} = -B_{K^- pp}
- {\rm i} \, \Gamma_{K^- pp} /2$, where $B_{K^- pp}$ is a binding energy
with respect to the $K^- pp$ threshold, $\Gamma_{K^- pp}$ is a width
of a quasi-bound state. The dependence of the
real and imaginary parts of the three-body pole energy as function of the
range parameter $\beta$ is shown in Figs.~\ref{Re_diffbetas.fig}
and~\ref{Im_diffbetas.fig}, respectively. It is
seen that the dependence of the real part on $\beta$ is rather weak,
whereas the imaginary part strongly depends on this parameter.

Other values which are varied are the mass $M_{\Lambda}$ and the width
$\Gamma_{\Lambda}$ of the $\Lambda(1405)$ resonance.
The results of such variations are shown in
Table~\ref{E_Lambda.tab}. All other input data used in this calculation
are fixed at $\beta=3.5$ fm$^{-1}$ and $a_{K^- p}^{\rm KEK}$.
As expected, the broadening of $\Lambda(1405)$ leads to a considerable
increase of the three-body width, whereas the three-body binding energy
depends on $\Gamma_{\Lambda}$ rather weakly. However, increasing the
$\Lambda(1405)$--resonance mass strongly affects both real and imaginary
parts of the three-body pole, leading to a fast decrease of both.
\begin{table}
\caption{Calculated three-body pole energy $E_{K^- pp}$ in MeV, of the
$I=1/2,~J^{\pi}=0^-$ quasi-bound state of the $\bar{K} NN$ system
with respect to the $K^-pp$ threshold, for different
two-body input, mass $M_{\Lambda}$ and half-width $\Gamma_{\Lambda}/2$
of the $\Lambda(1405)$. For $E_{\Lambda}=1420-{\rm i}~20$ MeV no
reasonable $T_{\bar{K}N-\pi \Sigma}$ parameters can be found. }
\label{E_Lambda.tab}
\begin{tabular}{cccc}
\hline \hline
$\Gamma_{\Lambda}/2$ $\setminus$ $M_{\Lambda}$ \, & \, 1400 \, &
\, 1410 \, & \, 1420 \, \\
\hline
\, $20$  \, & \, $-62.1-{\rm i}~46.9\, $ & \, $-47.5-{\rm i}~37.6$ \, &
\, {\it no $T_{\bar{K}N-\pi \Sigma}$} \, \\
\, $25$  \, &\, $-64.9-{\rm i}~58.4$ \,& \, $-50.8-{\rm i}~47.4$ \, &
\, $-40.6-{\rm i}~39.4$ \, \\
\, $30$  \, & $-65.7-{\rm i}~72.2$ & $-52.5-{\rm i}~59.8$ &
$-42.8-{\rm i}~50.8$ \\
\hline \hline
\end{tabular}
\end{table}

\subsection{One-channel real and complex $\bar{K}NN$ calculations}

We also performed a test calculation for the one-channel
$\bar{K}NN$ system using a one-channel real $\bar{K}N$
potential ($T$-matrix). For fitting we used the real part of
$a_{K^- p}^{\rm KEK}$, the real part of $E_{\Lambda}^{\, \rm PDG}$, and assumed
$\Lambda(1405)$ as a real bound state of the $I=0$ $\bar{K}N$ subsystem.
For these data, and using $\beta=3.5$ fm$^{-1}$, we found
a real bound state for $I=1/2$, $J^{\pi}=0^-$ $\bar{K}NN$ at $-43.8$~MeV below
the $K^- pp$ threshold (the first column in Table~\ref{E_complex.tab}).
\begin{table}
\caption{Results of different calculations of the three-body pole
energy $E_{K^- pp}$ in MeV, with respect to the $K^-pp$ threshold: real and complex
$\bar{K}NN$ one-channel (first two columns), and full coupled-channel
calculations (third column) using the `best set' of $\bar{K}N - \pi \Sigma$
parameters. Fourth column: complex $\bar{K}NN$ one-channel calculation with
`AY set'. Fifth column: AY's result~\cite{Akaishi2}. }
\label{E_complex.tab}
\begin{tabular}{ccccc}
\hline \hline
\, $E_{1 \,\rm real}^{\, \rm best}$ \, & \, $E_{1 \,\rm complex}^{\, \rm best}$ \, & \,
$E_{2 \,\rm coupled}^{\, \rm best}$ \, & \, $E_{1 \,\rm complex}^{\,\rm AY}$ \, &
\, $E$ from Ref.~\cite{Akaishi2} \, \\
\hline
\, $-43.8$  \,  & \, $-40.2-{\rm i}~38.7$ \, &
\, $-55.1-{\rm i}~50.9$ \, & \, $-46.6-{\rm i}~29.6$ \, &
\, $-48.0-{\rm i}~30.5$ \, \\
\hline \hline
\end{tabular}
\end{table}

\begin{figure}[h]
\includegraphics[scale=0.4,angle=-90]{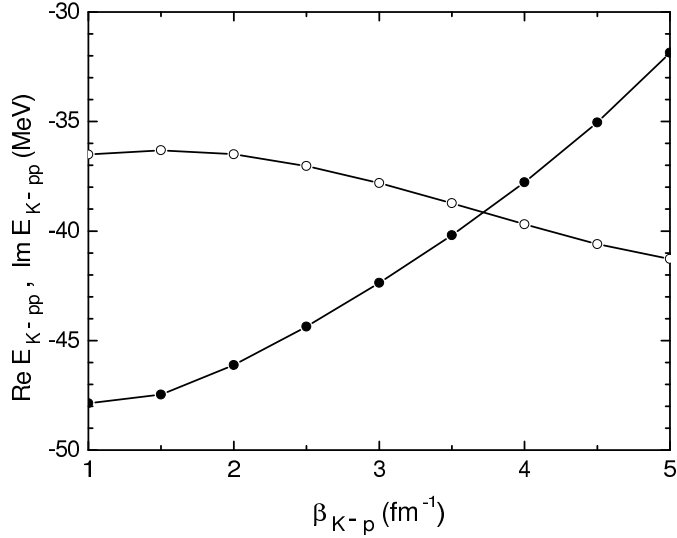}
\caption{\label{diff_betas_cmplxKN.fig} One-channel calculation
with complex $\bar{K}N$ potential:
the dependence of the real (solid circles) and imaginary (open circles)
parts of three-body $\bar{K}NN$ pole energy on the
$\bar{K}N$ range parameter $\beta$. The two-body $\bar{K}N$
observables are fixed at $a_{K^- p}^{\rm KEK}$ and $E_{\Lambda}^{\, \rm PDG}$.}
\end{figure}

Another test calculation was performed with a one-channel
complex $\bar{K}N$ potential.
The strength parameters $\lambda$ of the potential were fitted to
the $a_{K^- p}^{\rm KEK}$ and $E_{\Lambda}^{\, \rm PDG}$ data, and the
dependence of the three-body pole on the range parameter $\beta$
was investigated. Results are presented in
Fig.~\ref{diff_betas_cmplxKN.fig}.
\begin{figure}
\includegraphics[scale=0.4,angle=-90]{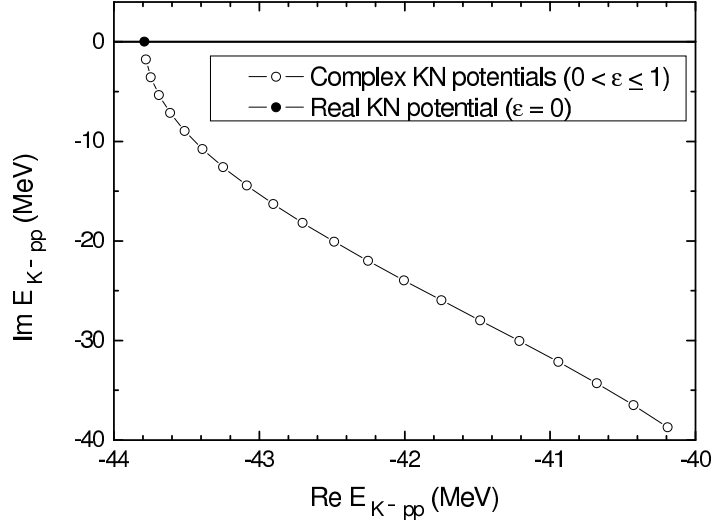}
\caption{\label{epsilonKN_cmplxKN.fig}
Trajectory of the three-body pole in the complex energy plane
from $E_{1 \,\rm real}^{\, \rm best}$, corresponding to a real
$\bar{K}N$ potential with $\varepsilon = 0$ (solid point), to
$E_{1 \,\rm complex}^{\, \rm best}$ for a complex $\bar{K}N$
potential with $\varepsilon = 1$ (see text for details).}
\end{figure}

It is seen from the plot that increasing the range of the $\bar{K}N$
interaction, by decreasing the range parameter $\beta$, gives rise to
a deeper and somewhat narrower three-body level.
The dependence of the calculated $\bar{K}NN$ energy on the range parameter
$\beta$, as displayed in Fig.~\ref{diff_betas_cmplxKN.fig}, is rather
strong. Therefore, using a too large or a too small range parameter
for the complex $\bar{K}N$ interaction leads to substantial underestimate
or overestimate, respectively, of the three-body energy.
The `best set' of $\bar{K}N$ parameters with a fixed value
for the range parameter, $\beta = 3.5$~fm$^{-1}$, yields the three-body
pole energy $E_{1 \,\rm complex}^{\,\rm best}$ shown in the second column of
Table~\ref{E_complex.tab}. The result of the full coupled-channel
calculation $E_{2 \,\rm coupled}^{\,\rm best}$ is shown in the third column.

The transition within a three-body single-channel $\bar{K}NN$ calculation
from using a real $\bar{K}N$ interaction to using the
complex $\bar{K}N$ interaction, fitted to $E_{\Lambda}^{\, \rm PDG}$ and
to $a_{K^- p}^{\rm KEK}$, is demonstrated in Fig.~\ref{epsilonKN_cmplxKN.fig}
by the trajectory of complex three-body energies starting with the real
$E_{1 \,\rm real}^{\,\rm best}$ at the upper-left corner and ending with the complex
$E_{1 \,\rm complex}^{\,\rm best}$ at the lower-right corner. This trajectory is
generated by varying a real parameter $\varepsilon$ between 0 to 1,
$\varepsilon = 0$ for $E_{1 \,\rm real}^{\,\rm best}$ and $\varepsilon = 1$ for
$E_{1 \,\rm complex}^{\,\rm best}$, such that the imaginary parts of the fitted
$E_{\Lambda}^{\, \rm PDG}$ and $a_{K^- p}^{\rm KEK}$ are scaled down by $\varepsilon$:
\begin{equation}
{\rm Im \,}{E_{\Lambda}^{\, \rm PDG}} \to \varepsilon \,{\rm Im \,}
{E_{\Lambda}^{\, \rm PDG}} \,, \quad
{\rm Im \,} a_{K^- p}^{\rm KEK} \to \varepsilon \,{\rm Im \,}{a_{K^- p}^{\rm KEK}} \,.
\end{equation}

It is interesting to note that although the $I=0$ and $I=1$ strength
parameters $\lambda_{\rm complex}$ provide stronger attraction in the
$\bar{K}N$ systems than the attraction provided by $\lambda_{\rm real}$,
yet $E_{1 \,\rm complex}$ signifies less binding than $E_{1 \,\rm real}$.
This generalizes the well known property in two-body problems where
including absorptivity leads effectively to adding repulsion.
Here we find that absorption of flux from the $\bar{K}N$ channel into other
unspecified channels represented by an imaginary part of a complex
$\bar{K}N$ potential reduces also the three-body binding energy.

Comparing the result of the one-channel complex $\bar{K}NN$
calculation with the coupled-channel $\bar{K}NN$
(see Table~\ref{E_complex.tab}) shows that $E_{2 \,\rm coupled}$ is much deeper
and broader than $E_{1 \,\rm complex}$. This means that the $\pi \Sigma$ channel,
within a genuinely three-body coupled-channel calculation plays an
important dynamical role in forming the three-body resonance
(quasi-bound state), over its obvious role of absorbing flux from the
$\bar{K}N$ channel.
The poor applicability of an optical potential approach (or some
low-order perturbation calculation) in searching for a quasi-bound state
was shown, for example, by Ueda~\cite{Ueda}, who studied the
$\eta NN - \pi NN$ coupled-channel system using Faddeev equations,
finding a large deviation of the calculated results from optical-model
predictions.

In order to compare the present results with the results of calculations
by Yamazaki and Akaishi~\cite{Akaishi2}, the one-channel $\bar{K}NN$
calculation was repeated using the complex $\bar{K}N$ potential
corresponding to the `AY set' of $\bar{K}N$ parameters.
The result obtained by us ($E_{1 \,\rm complex}^{\,\rm AY}$) and $E$
from Ref.~\cite{Akaishi2} are shown in the last two columns of
Table~\ref{E_complex.tab}. It is remarkable that in spite of different
forms of the two-body potentials and different three-body formalisms,
the calculated three-body energies in these single-channel $\bar{K}NN$
calculations come out very close to each other, provided the same set
of $\bar{K}N$ parameters is fitted to. Nevertheless, both
values of three-body energy are far away from the three-body energy
of the complete coupled-channel calculation. One of the reasons
is the use of a complex $\bar{K}N$ potential in the single-channel
$\bar{K}NN$ calculations, another reason is the too small value,
$\beta = 1.5$~fm$^{-1}$, for the range
parameter used in these approximate calculations.

\section{Conclusion}

We performed coupled-channel few-body calculations of the
$I=1/2$, $J^{\pi}=0^-$ $\bar{K}NN$ system, finding a deeply bound and
broad quasi-bound state, which is a resonance in the $\pi \Sigma N$ channel.
The calculations yielded binding energy $B_{K^-pp} \sim 50 -70$~MeV and
width $\Gamma_{K^-pp} \sim 100$~MeV, in agreement with our earlier
results~\cite{ourPRL}. It was shown that
the explicit inclusion of the second channel is crucial for this system.
The dependence of the three-body energy pole on different forms and
parameters of the $\bar{K}N$ interaction, and on different ways of
reproducing $\bar{K}N - \pi \Sigma$ observables, was studied.
Most of these dependencies were found to be strong. In particular, it was
shown that a complex $\bar{K}N$ potential gives much shallower and
narrower three-body quasi-bound state than the full coupled-channel
calculation, which has the same range parameter and reproduces the
same $\bar{K}N - \pi \Sigma$ observables.

We compared our results with those of Yamazaki and Akaishi~\cite{Akaishi2},
demonstrating the shortcomings of these single-channel $\bar{K}NN$
calculations. Two more calculations of the same system appeared recently.
Dote and Weise~\cite{DoteWeise} have presented preliminary results of
a variational Anitsymmetrized Molecular Dynamics calculation for the
$K^- pp$ system within a single-channel $\bar{K}NN$ framework.
Their calculation focuses attention to the dependence of the calculated
real three-body binding energy on the range parameter of the Gaussian
$\bar{K}N$ interaction used. It includes perturbatively also a $p$-wave
$\bar{K}N$ interaction. Whereas a direct comparison between our
coupled-channel calculations and these single-channel calculations cannot
be made, the general criticism expressed above of the use of
a single-channel formalism applies also to this work.

A coupled-channel $\bar{K}NN - \pi \Sigma N$ calculation of the same
$K^- pp$ system was performed recently by Ikeda and Sato~\cite{IkedaSato}
with less emphasis on reproducing low-energy $\bar{K}N$ data.
The obtained binding energies are in a similar range to those presented
here, while the widths are consistently lower than those calculated in
the present work.

It is worthwhile to note that all the theoretical calculations
discussed above, including the present calculations, obtain binding energies
which are considerably below the binding energy $\approx 115$~MeV deduced for
the $K^-pp$ identification proposed in Ref.~\cite{FINUDA}.
This FINUDA $K^-_{\rm stop}$ experiment on lithium and heavier targets,
as mentioned in the Introduction, leaves room for other interpretations
as well. The use of a more restrictive $^3$He target in order to search
for a $K^-pp$ quasibound state in a ($K^-,n$) reaction was approved as a `day-1'
experiment in J-PARC \cite{Nagae}.
The spectrum calculated recently for this reaction \cite{Koike} demonstrates
how the large width predicted
for $K^-pp$ in the present work is expected to wipe out any clear peak structure
in this reaction.

Additional calculations are necessary to study other features of
the coupled $\bar{K}NN$ system. These include the secondary effect of the
$\pi \Lambda$ channel beyond that of the primary inelastic $\pi \Sigma$
channel incorporated here, of $p$-wave $\bar{K}N$ and $\pi N$ interactions,
and the use of relativistic kinematics. Finally, in order to understand
better the $\bar{K}N$ interaction, it is desirable to
perform coupled-channel Faddeev calculations of a quasi-bound state in
the $S=1$ $\bar{K}NN$ system as well.

\vspace*{-5mm}
\begin{acknowledgments}
\vspace*{-5mm}
The work was supported by the Czech GA AVCR grant A100480617 and by the
Israel Science Foundation grant 757/05.
\end{acknowledgments}


\end{document}